\documentstyle[aps,multicol]{revtex}

\begin{document}
\bibliographystyle{myprb1}
\title{Granulated superconductors:
from the nonlinear $\sigma$ model to the Bose-Hubbard description 
}
\author{I. V. Yurkevich, and Igor V. Lerner}
\address{School of Physics and Astronomy, University of Birmingham,
Edgbaston, Birmingham B15 2TT, UK}

\date{July 20, 2000}
\maketitle

\begin{abstract}
We modify a nonlinear $\sigma$ model (NL$\sigma$M) for the description of a
granulated disordered system in the presence of both the Coulomb repulsion
and the Cooper pairing. We show that under certain controlled approximations
this model is reduced to 
the Bose-Hubbard (or ``dirty-boson'')
model with renormalized coupling constants. We obtain a more general
effective action (which is still simpler than the full NL$\sigma$M action)
which  can be applied in the region of parameters where the reduction to the 
Bose-Hubbard model is not justified. This action
may lead to a different picture of
the superconductor-insulator transition in 2D systems.
\end{abstract}

\begin{multicols}{2}
 A wide variety of experimental data on the super\-conductor--insulator 
(SI) transition
in two-dimen\-sional structures \cite{MK:99,SIT:99a,SIT:99b,GM:98}
continues to attract acute 
theoretical interest. The transition can be tuned by either disorder (changing
with the thickness of a superconducting film) or magnetic field,
thus being one of the most intensely studied examples of quantum phase 
transitions \cite{QPT:97}. 
However, recent experiments\cite{MK:99,SIT:99a} have challenged the very
existence of the SI transition, leaving open a possibility that a dramatic drop 
in resistance is due to the existence of a crossover to a new 
metallic phase with resistance much lower than that in 
the normal state and to a subsequent metal-superconducting
transition\cite{MK:99}. 
This situation requires a reassessment of theoretical approaches 
to the problem of dirty superconductors.

One of the ways to understand the problem of the SI transition is based on
the so-called Bose-Hubbard (or ``dirty-boson'') models
\cite{FGG:90}
where the superconducting 
phase is due to the bose-condensation of charge-$2e$ bosons (preformed Cooper
pairs) with localized vortices while the insulating phase is due to the
bose-condensation of 
vortices with localized Cooper pairs.
Another approach which captures the basic physics of granular superconductors 
is based on dissipative models\cite{CL:81}
of resistively shunted charged Josephson arrays\cite{AES:82,LO:83,AES:84,%
MPAF:86}, with the emphasis on the role of dissipation and Coulomb interaction.
In both group of models \cite{FGG:90,CL:81,AES:82,LO:83,AES:84,%
MPAF:86}, the transition is driven by fluctuations of the phase of the order
parameter.
An alternative approach is based on a 
microscopic description of {\it homogeneous} systems that incorporates both
the attractive (in the Cooper channel) and repulsive
electron-electron interaction in the presence of disorder 
into an effective field theory, the  nonlinear $\sigma$ model (NL$\sigma$M)
\cite{Fin:87,Fin:94,BeKi,FLS:00}. The SI transition in these models
is driven by fluctuations of the amplitude rather than the
phase of the order parameter, and the
Cooper pairing is suppressed by the repulsive interaction 
on the insulating side of the transition.

The purpose of this paper is to show that both the Bose-Hubbard model
\cite{FGG:90} and 
the dissipative models \cite{AES:82,LO:83,AES:84,%
MPAF:86} can be in fact derived from the
same microscopic action of a  NL$\sigma$M modified for the description of 
{\it granular} superconductors.
The two models correspond to certain simplifications made within the 
NL$\sigma$M. The latter is more general and allows one to go beyond different
limitations inevitable in the derivation of the   Bose-Hubbard
and dissipative models.
Note that the dissipative action\cite{AES:82} of 
Ambegaokar, Eckern, and Sch\"{o}n (AES) 
has been widely used\cite{SZ} in a simplified form in
the context of a normal tunnel
junction. This variant of the AES action
 has been very recently derived \cite{BEAH:00} from the NL$\sigma$M describing
electrons with the repulsive interaction moving in the presence of disorder. 
Here we will derive both the full AES action for Josephson junctions and the 
 Bose-Hubbard model from the  NL$\sigma$M that includes both the attraction in 
the Cooper channel and the Coulomb repulsion. We shall use a new variant 
\cite{YL:00} of the  NL$\sigma$M which, in our opinion, considerably simplifies
the calculations. Naturally, one can use (after a straightforward 
modification for a granular system) any version of the 
NL$\sigma$M that includes the Cooper pairing and the Coulomb interaction, 
either the original Finkelstein model\cite{Fin:87} or a more recent
 model\cite{FLS:00} in Keldysh technique.

Our starting point is the standard microscopic Hamiltonian which includes a
$\delta$-correlated Gaussian random potential, the Coulomb interaction and 
the BCS attraction. We consider a coarse-grained version of this Hamiltonian
which corresponds to a granular superconductor.
This will allow us  to separate scales of fluctuations of the amplitude and 
the phase of the
superconducting order parameter $\Delta$.
Neglecting (at some later stage) the amplitude
fluctuations and making some further simplifications
will lead eventually to the 
models\cite{FGG:90,AES:82} governed only by the phase fluctuations. 

Let us stress again that the NL$\sigma$M
used here  can be applied to the simultaneous 
description of both the amplitude and phase fluctuations of $\Delta$ 
that can be quite important in the relation to recent experiments
\cite{SIT:99a}. Moreover, even when focusing on the phase fluctuations only, 
the effective functional is essentially generalized by disorder (affecting
intra-granular electron motion and thus leading to a different model
of the phase fluctuations) and can in principle
lead to a different picture of the transition.

The derivation of the NL$\sigma$M from the Hamiltonian described above
follows the standard steps \cite{Fin:90}. First  one averages 
the replicated imaginary-time fermionic action over the random potential.
Then one employs the Hubbard-Stratonovich transformation to decouple
three quartic (in the 
electron field) terms, corresponding to the disorder, the Coulomb repulsion 
and the BCS attraction. Finally, integrating out the fermionic fields, one
arrives at the effective action in terms of three bosonic fields:
a matrix field $\hat \sigma$ which decouples the disorder-induced
``interaction'', $\Phi$ which decouples the Coulomb repulsion, and 
$\Delta$ which decouples the BCS attraction:
\begin{eqnarray}
{\cal S}[\hat\sigma, \hat\Delta,\Phi]=
\frac{\pi\nu}{8 \tau_{\text{el}}}\,{\mathrm Tr}\, \sigma ^2+
\frac1{4\lambda_{{\text{0}}}}{\mathrm Tr}\, |\hat\Delta|^2+\frac12 
{\mathrm Tr}\,\Phi U^{-1} \Phi
\nonumber
\\[-1mm]
\label{PDS}
\\[-1mm]
\nonumber
-\frac12 {\mathrm Tr}\ln\left[
-{\hat{\xi}}-\hat t
+\frac{i}{2 \tau_{\text{el}}}\,\hat\sigma+ i\left(
{\hat \Delta}+\Phi +\hat  \epsilon \right )\right]\,.
\end{eqnarray}
Here the operator $\hat \epsilon$ equals $i
\hat \tau_3 \partial _\tau$ in imaginary time representation 
and becomes the diagonal matrix of fermionic Matsubara frequencies in 
frequency representation, $\hat \xi$ is the operator of the intra-grain
kinetic energy (counted from the chemical potential), 
and $\hat t$ is the tunneling amplitude matrix (i.e.\ the inter-grain 
kinetic energy).
All the bosonic fields are defined in the space
which is convenient to think of as a
direct product of the $N\times N$ replica sector, $2\times 2$ spin sector,
$2\times 2$ `time-reversal' sector (introduced for a correct decoupling 
in the Cooper channel for both disorder-induced and BCS interaction), and 
of the $m\times m$ grain sector. The symbol Tr refers both to a summation over
all these matrix indices and to an integration over position 
${\bf r}$ and the imaginary time $\tau$. The matrix $\hat \sigma$ is diagonal 
in grain indices (called later $i,j$) and
possesses standard symmetries in all the other sectors\cite{Fin:90}.  
The pairing field $\hat \Delta$ is diagonal in
the replica and grain indices and in $x\equiv ({\bf r}, \tau)$, 
and has the following structure\cite{YL:00}
 in time-reversal and spin sectors:
\begin{equation}
\label{Delta}
{\hat \Delta}(x)= |\Delta(x)| e^{\frac{i}{2}\chi(x)\hat\tau_3}
\hat \tau^{sp}_2\otimes\hat\tau_2
e^{-\frac{i}{2}\chi(x)\hat\tau_3}\,,
\end{equation}
where $\hat \tau_\alpha$ and $\hat \tau^{\text{sp}}_\alpha$ 
are Pauli matrices in the time-reversal and spin sectors, respectively. 
The Coulomb field $\Phi(x)$ is proportional to the unit matrix in 
all the matrix sectors. Finally, in Eq.~(\ref{PDS})
$\lambda_{{\text{0}}}$ and $\tau_{\text{el}}$ are the BCS coupling 
constant and the elastic mean free time, respectively, and $U\equiv
U({\bf r}- {\bf r}')$ is the Coulomb interaction.

The principal simplification for granular systems is that
all the fields are spatially homogeneous inside each 
grain when the grains are zero-dimensional, i.e.\ their
 sizes $L\alt  \xi, L_T $ ($\xi $ and $L_T$ are
 the superconducting and thermal coherence lengths)
which is equivalent to $|\Delta|, T\alt 1/\tau_{\text{erg}}$.
Then the Coulomb interaction reduces to the
capacitance matrix, $U^{-1}\!\to \!C_{ij}/e^2$, and the  
tunneling matrix $\hat t =\{t_{ij}\}$ depends only on grain 
indices.

Now we follow the procedure of derivation of the NL$\sigma$M for dirty
superconductors \cite{YL:00}. First, we look for
a saddle point of (\ref{PDS})
with respect to $\hat \sigma$ separately for each grain. 
It can be parameterized as $\hat \sigma_{\text{s.p.}}\!=\!S^\dagger \Lambda
S$, where $S$ is a certain matrix which also diagonalizes $\hat \epsilon
\!+\!\Phi \!+\!\hat \Delta$,
\begin{equation}
\label{diag}
{\hat \epsilon}_i+{\Phi}_i+{\hat \Delta}_i
=S_i^\dagger \lambda_iS_i\,,
\end{equation}
where all the matrices are diagonal in grain indices. Then the 
entire saddle-point manifold is parameterized as
\begin{equation}
\label{SP}
\hat \sigma_i=S_i^\dagger Q_i S_i, \qquad Q_i=U_i^\dagger \Lambda U_i\,,
\end{equation}
where in the Matsubara representation $\Lambda
={\text {diag\,\{sgn\,$\epsilon$\}}}$, and  matrix $U$ defines
the standard coset space\cite{EfLKh}.
Here $Q$ is a degenerate solution to the saddle-point equation for the 
action (\ref{PDS}) when ${\hat \epsilon}=0$,
$\hat \Delta $ and $\Phi$ all vanish.
 The parameterization (\ref{SP}) `aligns' the field
$\sigma$ so that $\hat \Delta $ and $\Phi$ are taken into account 
in the zeroth approximation. 

Now we perform a similarity transformation with matrices $S$ and $S^\dagger $
under Tr\,ln in Eq.~(\ref{SP}). As all the fields are spatially homogeneous
inside each grain, $S$ commutes with the operator $\hat \xi$. Then one only 
needs to expand the Tr\,ln to the first nonvanishing orders in $t_{ij}$ and
$\lambda_i$, this expansion being justified when $|t|, |\Delta|, T\ll
1/\tau_{\text{el}}\ll \varepsilon_{\text{\sc f}}$. Thus one arrives
 at the following effective action: 
\begin{eqnarray}
{\cal S}[Q,\Delta,\Phi]=
\int\limits_0^\beta{\rm d}\tau \biggl\{
\sum_i\frac{|\Delta_i|^2}{\nu\lambda_0\delta_i}
+
\sum_{ij}\frac{C_{ij}}{2e^2}\,\Phi_i\Phi_j
\biggr\}\nonumber
\\[-2mm]
\label{action}
\\[-2mm]
\nonumber
-
\sum_i\frac{\pi}{2\delta_i}{\mathrm Tr}\,\lambda_iQ_i -
\frac{g_{ij}^{\text{\sc t}}}2
\sum_{ij}{\mathrm Tr}\,Q_iS_{ij} Q_j S_{ji} \,,\quad
\end{eqnarray}
where $S_{ij}\equiv S_iS_j^\dagger$, all the fields depend on $\tau$,
Tr refers  to all indices except those
numerating grains, $\delta_i$ is mean level
spacing in the $i$-th grain, and the tunneling
conductance is defined by $g_{ij}^{\text{\sc t}}\equiv
2\pi^2|t_{ij}|^2/\delta_i \delta_j$ (which is nonzero only for
neighboring grains).
Both $S$ and $\lambda$ should be found from the diagonalization procedure in 
Eq.~(\ref{diag}).

The next step is to represent $S_i$ as 
\begin{equation}
\label{SV}
S_i=V_i\, {\rm e}^{-\frac i2 \chi_i(\tau)\,\hat \tau_3}\,.
\end{equation}
This is similar to the gauge transformation suggested in
Refs.~\cite{KamGef:96}
and used in Ref.~\cite{BEAH:00} to gauge out the Coulomb field. 
However, one  cannot gauge out two independent fields, $\Delta $ and $\Phi$.
Substituting the transformation (\ref{SV}) into the diagonalization condition
(\ref{diag}), we reduce it to 
\begin{equation}
\label{diag2}
{\hat \epsilon}+\widetilde\Phi_i
+{\hat \Delta}_i^{\!\text o}
=V_i^+\lambda_i V_i\,,
\end{equation} 
where ${\hat \Delta}_i^{\!\text o}$ is the field (\ref{Delta}) taken at $\chi
=0$ and the field $\widetilde\Phi$ is given  in the $\tau$ representation by
$\widetilde\Phi_i\equiv
{\Phi}_i- \case 12 \partial _\tau \chi_i
\,.
$

Both $ \widetilde\Phi $ and $|\Delta |$ are massive fields 
whose fluctuations are strongly suppressed.
It is straightforward to show that the fluctuations of $ \widetilde\Phi $ 
are of order $\delta$ which is much smaller than both $T$ and $|\Delta|$.
 Therefore, in the mean field 
approximation  in $ \widetilde\Phi $  this field can be 
neglected,  $ \widetilde\Phi=0$.
This condition is nothing more than the Josephson relation in imaginary
time\cite{LO:83}. This
locks the fluctuations of the Coulomb field $ \Phi$ with 
the phase-fluctuations of the pairing field $\hat \Delta$, 
\begin{eqnarray}
\label{Phi}
\Phi=\case12\partial _\tau \chi\,,
\end{eqnarray}
 thus reducing  the action (\ref{action})
 to one depending only on the fields $Q$ 
and $\Delta$.   

The mean field approximation  in $|\Delta_i|$ is valid
for $|\Delta_i|\gg\delta$ and and reduces to
the standard self-consistency equation  which formally follows from the 
variation of the action (\ref{action}) with respect to $|\Delta|$. In this
approximation one finds $|\Delta_i|$ to be independent of $i$ and $\tau$.
Thus the first term in Eq.\ (\ref{action}) becomes a trivial constant, 
so that the action depends on  $Q$ and $\chi$ only:
\begin{eqnarray}
{\cal S}[Q,\chi]&=&
\sum_{ij}\frac{C_{ij}}{8e^2}
\int\limits_0^\beta\!\!{\rm d}\tau\,
\partial _\tau \!\chi_i\,\partial _\tau \!\chi_j
- \sum_i\frac{\pi}{2\delta_i}{\mathrm Tr}\,\lambda_i Q_i -
\nonumber
\\[-2mm]
\label{action2}
\\[-2mm]
\nonumber
&-&\frac{g_{ij}^{\text{\sc t}}}2
\sum_{ij}{\mathrm Tr}\,Q_iS_{ij}Q_jS_{ji}\,.
\end{eqnarray}
Note that the field $\chi $ in this action obeys the standard boundary 
condition $\chi(\tau +\beta)= \chi(\tau)\,\text{mod}\,2\pi$.
Thus when calculating
the partition function with this action, one should take into account different
topological sectors corresponding to different winding numbers in $\chi$.

 Now the diagonalization conditions (\ref{diag2})
are the same for each grain, and 
reduce to those solved in Ref.~\cite{YL:00}:
\begin{eqnarray}
V_{\epsilon\epsilon'}
= \cos\frac{\theta_\epsilon}2\,\delta_{\epsilon,\epsilon'}
+
\hat \tau^{sp}_2\!\otimes\!\hat\tau_2\,
\sin\frac{\theta_\epsilon}{2}\,{\rm sgn}\epsilon\,\delta_{\epsilon,-\epsilon'}
\nonumber
\\[-1mm]
\label{diagonal}
\\[-1mm]
\nonumber
\lambda={\mathrm diag}
\sqrt{\epsilon^2   + |\Delta|^2}\,{\mathrm sgn}\, \epsilon \,,\quad
\cos\theta_\epsilon\equiv
\frac{|\epsilon|}{\sqrt{\epsilon^2   + |\Delta|^2}}
\,.
\end{eqnarray}
Then $S_{ij}$ in Eq.~(\ref{action2})
can be expressed in terms of  $V$ as 
\begin{eqnarray}
\label{Sij}
S_{ij} \equiv V
{\rm e}^{-\frac i2 \chi_{ij}\,\hat \tau_3}V^\dagger\,,
\qquad \chi_{ij}\equiv  \chi_{i} -  \chi_{j}\,.
\end{eqnarray}
Finally note that large-$|\epsilon|$ contributions to the action (\ref{SP})
are strongly suppressed, while for $|\epsilon|\ll |\Delta|$ 
one has $\lambda =|\Delta| \Lambda$ which suppresses
fluctuations of
$Q$ in each grain imposing $Q\!=\!\Lambda$.
Then, all matrices in the action (\ref{action2}) are diagonal
in the replica indices so that these indices become redundant. The 
diagonalization procedure (\ref{diagonal})--(\ref{Sij}) has resolved explicitly
the matrix dependence on the time-reversal and spin indices.
This reduces the action to that depending only on one {\it scalar} bosonic
field, the phase $\chi $ of the order parameter, whose arguments are the 
imaginary time $\tau$, and the position index $i$, i.e.\ the grain
number. Indeed, the second term in Eq.~(\ref{action2}) reduces to a trivial
constant;  evaluating the tunneling term with the help of Eqs.~(\ref{diagonal})
and (\ref{Sij}), we obtain
\begin{eqnarray}
\label{tunnel}
\lefteqn{\qquad{\cal S}[\chi]=
\sum_{ij}\biggl\{\frac{C_{ij}}{8e^2}
\int\limits_0^\beta\!\!{\rm d}\tau\,
\partial _\tau \!\chi_i\,\partial _\tau \!\chi_j
}\\
&&-2{g_{ij}^{\text{\sc t}}}
\int\limits_0^\beta\!\!{\rm d}\tau\!\!
 \int\limits_0^\beta\!\!{\rm d}\tau '
g_{\text n}^2 (\tau \!-\!\tau ') 
\cos \chi^-_{ij}+
g_{\text a}^2 (\tau \!-\!\tau ') 
\cos \chi^+_{ij}
\biggr\}\nonumber
,
\end{eqnarray}
where $\chi_{ij}(\tau)\equiv\chi_{i}(\tau)-\chi_{j}(\tau)$,
$$
 \chi_{ij}^\pm\equiv\case12 \Bigl[
\chi_{ij}(\tau)\pm \chi_{ij}(\tau')\Bigr]
\,,
$$
and the normal and anomalous Green's functions $g_{n,a}$
(integrated over all momenta)  are given by
\begin{eqnarray}
\label{GF}
g_n(\tau)\!=\!T\!\sum_{\epsilon}\!\frac{\epsilon\sin\epsilon\tau}
{\sqrt{ \epsilon^2 \!+\! |\Delta|^2}}\,,
\;
g_a(\tau)\!=\!T\!\sum_{\epsilon}\!\frac{|\Delta|\cos\epsilon\tau}
{\sqrt{ \epsilon^2 \!+\! |\Delta|^2}}\,.
\end{eqnarray}
This action coincides with that derived in
Refs.~\cite{AES:82,LO:83,AES:84}.
Further simplifications are possible in two limiting cases. 

First,  in the normal case ($\Delta\!=\!0$), one has in Eq.~(\ref{GF})
\begin{eqnarray}
\label{GF2}
g_a=0\,,\qquad g_n^2(\tau) =\frac{T^2}{\sin^2\pi T\tau}\,.
\end{eqnarray}
Then the field $\chi$ should be substituted, according to
Eq.~(\ref{Phi}), by
$2\int^\tau\!{\rm d}\tau' \Phi(\tau')$. This limiting case corresponds to
using the action (\ref{tunnel}), (\ref{GF2}) 
in the context of a normal tunnel junction\cite{SZ}.
This is precisely the action which has been recently derived from the 
NL$\sigma$M in Ref.~\cite{BEAH:00}; the functional (\ref{action2}) in the
limit $\Delta=0$ is equivalent to the $\sigma$ model of  Ref.~\cite{BEAH:00}.
Including the disorder-induced fluctuations (i.e.\ going
beyond the $Q=\Lambda$ approximation) allows one to obtain \cite{BEAH:00}
a correct low-$T$ limit for the phase correlation function missing in
the action (\ref{tunnel}).

The action (\ref{action2}) is more general than that considered in
Ref.~\cite{BEAH:00}: although
under the mode locking condition (\ref{Phi})  it depends
only on the fields $\chi$ and $Q$, 
the matrix $S_{ij}$, Eqs.~(\ref{diagonal}) and
(\ref{Sij}), reduces to a simple $U(1)$ gauge transformation as in
in  Ref.~\cite{BEAH:00} only in the limit $\Delta=0$.   

The second limiting case,
$T\ll |\Delta|$, is just the limit relevant in 
the context of the SI transition in granular superconductors. For $T=0$, the
summation in Eq.~(\ref{GF}) can be substituted by integration which yields
\begin{eqnarray*}
g_n(\tau)=\frac{|\Delta|}{\pi}\,K_1(|\Delta|\tau), \quad
g_a(\tau)=\frac{|\Delta|}{\pi}\,K_0(|\Delta|\tau)
\end{eqnarray*}
This is also a good approximation for a low-temperature case; substituting this
into Eq.\ (\ref{tunnel}) gives
the action for the dissipative model\cite{LO:83,AES:84}. 
Note that for $|\tau-\tau'|\ll|\Delta|^{-1}$, the main contribution in the
tunneling action (\ref{tunnel}) is given by the normal term with the
corresponding kernel proportional to $|\tau-\tau'|^{-2}$. The Fourier transform
of this would give a term of the Caldeira-Leggett type\cite{CL:81}
proportional to $|\omega|$. 

The tunneling action (\ref{tunnel}) is non-local in $\tau$. As has been noted
in Ref.~\cite{LO:83} for the case of one tunnel junction, 
for sufficiently large capacitance the phase $\chi_{ij}$ changes slowly in 
comparison with $|\Delta|^{-1}$, and in the adiabatic approximation $\chi(
\tau`)$ is changed by $\chi(\tau) +(\tau'\!-\!\tau)\partial_
\tau\chi(\tau)$. Making such an expansion, one obtains from Eq.~(\ref{tunnel}) 
the following local action:
\begin{eqnarray}
\label{DB}
{\cal S}[\chi]=
\int\limits_0^\beta\!\!{\rm d}\tau\,\biggl\{ 
\sum_{i j} \case12u^{-1}_{ij} \dot \chi_i \dot \chi_j
-|\Delta| {g_{ij}^{\text{\sc t}}}
\cos \chi_{ij}
\biggr\}
,
\end{eqnarray}
where $\dot \chi_i\equiv \partial _\tau \!\chi_i$ and 
\begin{eqnarray*}
\frac1{u_{ii}}&\equiv& \frac{C_{ii}}{4e^2} +\sum_j\frac{{g_{ij}^{\text{\sc t}}}}
{|\Delta|}\frac{3+\cos \chi_{ij}}8\,,\\
u^{-1}_{ij} &\equiv& \frac{C_{ij}}{4e^2} -\frac{{g_{ij}^{\text{\sc t}}}}
{|\Delta|}\frac{3+\cos \chi_{ij}}8\,.
\end{eqnarray*}
If all the self-capacitances are equal to $C$ with $E_c\propto e^2/C$ being the
charging energy, and all ${g_{ij}^{\text{\sc t}}}={g^{\text{\sc t}}}$,
then $u_{ii}\equiv U$ has the meaning of the 
renormalized charging energy. Ignoring 
 a weak dependence of $u$ on $\cos \chi_{ij}$
in  the above relations, one obtains the renormalized charging energy:
\begin{eqnarray}
\label{U}
U=\frac {E_c}{1+\# E_c {g^{\text{\sc t}}}/|\Delta|  }\,.
\end{eqnarray}
Here the coefficient $\#$ depends on the number of next neighbors for each
grain, etc. A similar renormalization takes place for the next-neighbor 
off-diagonal energy $u_{ij}$.
Now one can see that on the face of it the adiabatic approximation employed to 
obtain Eq.~(\ref{DB}) is valid for $U\ll |\Delta| $. However, in the region
$g^{\text{\sc t}}\gg |\Delta|/E_c$, 
where the charging energy (\ref{U}) is strongly renormalized, 
the instanton-like solutions \cite{Korshunov} may be important.
This may further reduce the region of applicability for the local in $\tau$ 
action (\ref{DB}). 

Finally, by introducing the operator $\hat n$ canonically conjugate to the 
phase $\chi$, one finds the Hamiltonian that corresponds to the action 
(\ref{DB}):
\begin{eqnarray}
\label{HDB}
\hat H=
\sum_{i j} \case12u_{ij} \hat n _i \hat n_j
-|\Delta| {g_{ij}^{\text{\sc t}}}
\cos\left( \chi_{i}-\chi_j\right)\,
.
\end{eqnarray}
This is just the Hamiltonian of the Bose-Hubbard model\cite{FGG:90} which 
was first microscopically derived by Efetov \cite{Ef:80} in the context of 
granulated superconductors.

To conclude, we have shown that for a granular system with zero-dimensional 
grains the NL$\sigma$M reduces to the AES model (\ref{tunnel}).
When the charging energy is much smaller than
$|\Delta|$, the action (\ref{tunnel}) further reduces
to the Bose-Hubbard model, Eq.~(\ref{HDB}),
 which is widely used for the description of the superconductor-insulator
transition\cite{FGG:90}.
However, the above estimations show that this reduction is parametrically 
justified only for  the region $E_c\ll |\Delta|$ where the transition happens
at $g^{\text{\sc t}}\ll E_c/ |\Delta|\ll1$ which corresponds to a strongly 
granulated system.
Note finally that
 the most general (in the present context)  action (\ref{action})
describes both amplitude and phase fluctuations of the order parameter, 
being still considerably different from the NL$\sigma$M action for homogeneous
systems.
We hope that using this action may eventually lead to a different phase diagram 
for granulated superconductors.

\acknowledgments
This work has been supported by the Leverhulme  Trust under the contract
 F/94/BY.


\begin{thebibliography}{10}

\bibitem{MK:99}
N.~Mason and A.~Kapitulnik, {\it Phys. Rev. Lett.} {\bf 82}, 5341 (1999);
cond-mat/0006138 (2000).

\bibitem{SIT:99a}
J.~A. Chervenak and J.~M. Valles~Jr., {\it Phys. Rev. {\rm B}} {\bf 59}, 11209
  (1999).

\bibitem{SIT:99b}
N.~Markovic, C.~Christiansen, A.~M. Mack, W.~H. Huber, and A.~M. Goldman, {\it
  Phys. Rev. {\rm B}} {\bf 60}, 4320 (1999).

\bibitem{GM:98}
For a review of earlier experiments, see
A.~M. Goldman and N.~Markovic, {\it Physics Today} {\bf 51}, 39 (1998).

\bibitem{QPT:97}
For a review, see 
S.~L. Sondhi, S.~M. Girvin, J.~P. Carini, and D.~Shahar, {\it Rev. Mod. Phys.}
  {\bf 69}, 315 (1997).

\bibitem{FGG:90}
M.~P.~A. Fisher, G.~Grinstein, and S.~M. Girvin, {\it Phys. Rev. Lett.} {\bf
  64}, 587 (1990);
M.~P.~A. Fisher, {\it ibid} {\bf 65}, 923 (1990);
E.~S. Sorensen et al., {\it ibid} {\bf 69}, 828 (1992);
C.~Bruder, R.~Fazio, and G.~Sch\"{o}n, {\it Phys. Rev. {\rm B}} {\bf 47}, 342
  (1993);
M.~Wallin et al., {\it ibid} {\bf 49}, 12115 (1994);
J.~Lidmar  et al., {\it ibid} {\bf 58}, 2827 (1998);
D.~Das and D.~Doniach, {\it ibid} {\bf 60}, 1261 (1999).

\bibitem{CL:81}
A.~O. Caldeira and A.~J. Leggett, {\it Phys. Rev. Lett.} {\bf 46}, 211 (1981).

\bibitem{AES:82}
V.~Ambegaokar, U.~Eckern, and G.~Sch\"{o}n, {\it Phys. Rev. Lett.} {\bf 48},
  1745 (1982).

\bibitem{LO:83}
A.~I. Larkin and Y.~N. Ovchinnikov, {\it Phys. Rev. {\rm B}} {\bf 28}, 6281
  (1983).

\bibitem{AES:84}
U.~Eckern, G.~Sch\"{o}n, and V.~Ambegaokar, {\it Phys. Rev. {\rm B}} {\bf 30},
  6419 (1984).

\bibitem{MPAF:86}
S.~Chakravarty et al., {\it Phys. Rev. Lett.} 
 {\bf 56}, 2303 (1986);
M.~P.~A. Fisher, {{\it ibid} \bf 57}, 885 (1986);
D.~Dalidovich and P.~Phillips, {\it ibid} {\bf 84}, 737 (2000);
A.~Cuccolu, A.~Fibini and V.~Tognetti, {\it Phys. Rev. {\rm B}} {\bf 61},
11289 (2000).

\bibitem{Fin:87}
A.~M. Finkelshtein, {\it JETP Letters} {\bf 45}, 46 (1987).

\bibitem{Fin:94}
A.~M. Finkelstein, {\it Physica B} {\bf 197}, 636 (1994).

\bibitem{BeKi}
D.~Belitz and T.~R. Kirkpatrick, {\it Rev. Mod. Phys.} {\bf 66}, 261 (1994);
T.~R. Kirkpatrick and D.~Belitz, {\it Phys. Rev. Lett.} {\bf 79}, 3042 (1997).

\bibitem{FLS:00}
M.~V. Feigel'man, A.~I. Larkin, and M.~A. Skvortsov, {\it Phys. Rev. {\rm B}}
  {\bf 61}, 12361 (2000).

\bibitem{SZ}
G.~Sch\"{o}n and A.~D. Zaikin, {\it Phys. Rep.} {\bf 198}, 237 (1990).

\bibitem{BEAH:00}
I.~S. Beloborodov et al.,
  cond-mat/0006337 (2000).

\bibitem{YL:00}
I.~V. Yurkevich and I.~V. Lerner,
cond-mat/0006383 (2000).

\bibitem{Fin:90}
A.~M. Finkelstein, {\it Sov. Sci. Rev. 2} {\bf 14}, 1 (1990).

\bibitem{EfLKh}
K.~B. Efetov, A.~I. Larkin, and D.~E. Khmel'nitskii,  {\it Sov. Phys. JETP}
  {\bf 52}, 568 (1980).

\bibitem{KamGef:96}
A.~Kamenev and Y.~Gefen, {\it Phys. Rev. B} {\bf 54}, 5428 (1996).
A.~Kamenev and A.~Andreev, {\it Phys. Rev. {\rm B}} {\bf 60}, 3944 (1999).

\bibitem{Korshunov}
S.~E. Korshunov, {\it JETP Letters} {\bf 45}, 434 (1987);
G.~Sch\"{o}n and A.~D. Zaikin, {\it Phys. Rev. Lett.} {\bf 67}, 31 (1991);
X.~Wang and H.~Grabert, {\it Phys. Rev. {\rm B}} {\bf 53}, 12621 (1996).


\bibitem{Ef:80}
K.~B. Efetov,  {\it Sov. Phys. JETP} {\bf 51}, 1015 (1980).

\end{thebibliography}

\end{multicols} \end{document}